\documentclass[11pt]{article}
\usepackage{amsfonts}
\usepackage{amssymb}
\usepackage{mathrsfs}
\usepackage{graphicx}
\usepackage{amsmath,amsthm,amssymb,amscd}
\usepackage[all]{xy}
\usepackage{subfig}
\usepackage{hyperref}
\usepackage{multirow}
\usepackage{color}
\usepackage{float}
\usepackage{mathtools,slashed}
\usepackage{bbold}
\usepackage{bbm}
\usepackage{slashed}
\usepackage{tensor}

\usepackage{tikz}
\tikzstyle{process} = [rectangle, minimum width=3cm, minimum height=1cm, text centered, draw=black, fill=orange!30]
\tikzstyle{arrow} = [thick,->,>=stealth]

\usepackage{cite} 

\usepackage{amsxtra,graphics,epsfig,bm,tikz,xfrac,lscape}
\usetikzlibrary{decorations.pathmorphing}
\usetikzlibrary{decorations.markings}
\usetikzlibrary{arrows, decorations.markings, calc, fadings, decorations.pathreplacing, patterns, decorations.pathmorphing, positioning}

\usetikzlibrary{positioning,shapes}
\usetikzlibrary{chains}
\usetikzlibrary{arrows,fit,decorations.pathreplacing}
\tikzstyle{every picture}+=[remember picture]
\tikzstyle{na} = [baseline=-.5ex]

\newcommand{\nn}{\nonumber}

\def\eqa{\begin{eqnarray}}
\def\eqae{\end{eqnarray}}
\def\eq{\begin{equation}}
\def\eqe{\end{equation}}
\def\be{\begin{equation}}
\def\ee{\end{equation}}
\def\bea{\begin{eqnarray}}
\def\eea{\end{eqnarray}}
\def\ba{\begin{array}}
\def\ea{\end{array}}
\def\bd{\begin{displaymath}}
\def\ed{\end{displaymath}}

\def\>{\rangle}
\def\<{\langle}
\def\a{\alpha}
\def\b{\beta}

\def\g{\gamma}

\def\l{\lambda}
\def\m{\mu}
\def\n{\nu}

\def\r{\rho}
\def\s{\sigma}
\def\t{\theta}

\numberwithin{equation}{section}

\definecolor{darkblue}{rgb}{0,0,0.5}
\definecolor{darkred}{rgb}{0.5,0,0}
\definecolor{darkgreen}{rgb}{0,0.5,0}
\definecolor{orange}{rgb}{0.9,0.58,0}

\begin{document}

\begin{titlepage}
\hfill LCTP-20-22

\vskip 1 cm

\begin{center}
{\Large \bf Comments on Sen's Classical Entropy Function }\\

\vskip .7cm

{\Large \bf  for Static and Rotating AdS$_4$ Black Holes}\\

\end{center}

\vskip 1.4 cm
\begin{center}
{ \large \bf Jewel K. Ghosh${}^a$ and Leopoldo A. Pando Zayas${}^b$}
\end{center}

\vskip .4cm

\centerline{\it ${}^a$ International Centre for Theoretical Sciences, Tata Institute of Fundamental Research}
\centerline{\it  Shivakote, Bengaluru 560089, India}

\vskip .4cm \centerline{\it ${}^b$ Leinweber Center for Theoretical
Physics,   Randall Laboratory of Physics}
\centerline{ \it The University of
Michigan, Ann Arbor, MI 48109-1120}
\bigskip\bigskip

\centerline{\it ${}^b$ The Abdus Salam International Centre for Theoretical Physics}
\centerline{\it  Strada Costiera 11,  34014 Trieste, Italy}

\bigskip\bigskip


\vskip 1.5 cm
\begin{abstract}

\end{abstract}
We consider various aspects of Sen's classical entropy function  formalism for asymptotically AdS$_4$ black holes with  emphasis on its efficacy to capture higher derivative corrections to the Bekenstein-Hawking entropy.  The formalism has the important advantage of being based on near-horizon symmetries and does not require knowledge of the full interpolating supergravity solution,  nor of its AdS$_4$ asymptotics. For the static case, we focus on applying the entropy function formalism in the presence of various higher derivative terms  motivated in conformal supergravities; we find agreement with recently reported  results  utilizing  the full black hole solutions and Wald's entropy formula. For the rotating case, we demonstrate that a modified version of the formalism generates a background that coincides precisely with the Bardeen-Horowitz limit of known rotating, electrically charged AdS$_4$ black holes and provides a swift  approach to the black hole entropy, including higher derivatives corrections. We conclude that Sen's classical entropy function formalism is a viable and  highly efficient approach to capturing higher-derivative corrections to the entropy of asymptotically  AdS$_4$ black holes albeit naturally missing certain relations arising from global aspects of the full black hole solution. 

\vskip  1.5 cm

{\tt  jewel.ghosh@icts.res.in, \,lpandoz@umich.edu }
\end{titlepage}

\noindent
\section{Introduction}

The Bekenstein-Hawking entropy formula, stating that the thermodynamic entropy of black holes is a quarter of the horizon area in Planck units, is incredibly universal. It applies to black holes in asymptotically flat spacetimes as well as in asymptotically Anti-de-Sitter spacetimes; it is valid for static configurations as well as for rotating ones. This universality indicates that its origins are rooted   in general aspects of quantum gravity that are visible from the low energy regime. Therefore, to claim a deeper understanding of the Bekenstein-Hawking entropy formula one needs to turn to its corrections; only through the corrections we can definitely access the degrees of freedom underlying the formula.   In any theory of gravity there are two kinds of corrections to the Bekenstein-Hawking formula: (i) quantum corrections, related to virtual fields propagating in loops and (ii) higher curvature corrections. 
If we view the Einstein-Hilbert Lagrangian in the framework of an effective field theory expansion, it is simply the two-derivative term and will naturally be followed by higher-derivative terms. In this manuscript we focus on the study of higher-derivative corrections to the entropy of asymptotically AdS$_4$ black holes. 

The timing of our investigation stems from a number of important developments in our understanding of microscopic aspects of AdS black holes. The entropy of certain static, magnetically charged   AdS$_4$ black holes was provided a microscopic interpretation via the topologically twisted index of the dual field theory in \cite{Benini:2015eyy}  (see \cite{Zaffaroni:2019dhb}  for a review  and references).  More recently,  microscopic foundations  for rotating, electrically charged black holes in AdS$_5$ were provided via the superconformal index  \cite{Cabo-Bizet:2018ehj, Choi:2018hmj, Benini:2018ywd}; similar results have been extended to AdS$_{4,,6,7}$ \cite{Choi:2019miv,Choi:2019zpz, Kantor:2019lfo, Nahmgoong:2019hko, Nian:2019pxj,Bobev:2019zmz,Benini:2019dyp,Crichigno:2020ouj}. There have recently  been two studies exploring aspects of higher-derivative corrections for AdS black holes; one focused on AdS$_4$ solutions \cite{Bobev:2020egg} and another on AdS$_5$ \cite{Melo:2020amq}. Our approach to higher-derivative corrections  to the entropy of AdS$_4$ black holes is rooted in Sen's entropy function formalism introduced in  \cite{Sen:2005wa}.

Sen's entropy function formalism is powerful precisely because it does not require knowledge of the full  supergravity solution, nor of its AdS$_4$ asymptotics; it is  formulated in terms of symmetries of the near-horizon region    \cite{Sen:2005wa}.  This approach precisely allows us to study corrections to the entropy based only on the near-horizon geometry and complements the microscopic foundation to the entropy of AdS$_{4,5,6,7}$ rotating and electrically charged black holes  provided recently in \cite{David:2020ems} by using the Kerr-AdS/CFT correspondence \cite{Guica:2008mu,Lu:2008jk, Chow:2008dp}. The analysis of \cite{David:2020ems}, however,  started out from the higher dimensional solutions and applied the Bardeen-Horowitz near-horizon limit \cite{Bardeen:1999px} from which the algebra of asymptotic symmetries leads to a microscopic counting.

Let us further emphasize our big-picture question. The  AdS/CFT community is rightfully  elated by  the avalanche of microscopic entropy derivations \cite{Benini:2015eyy,Zaffaroni:2019dhb,Cabo-Bizet:2018ehj, Choi:2018hmj, Benini:2018ywd,Choi:2019miv,Choi:2019zpz, Kantor:2019lfo, Nahmgoong:2019hko, Nian:2019pxj,Bobev:2019zmz,Benini:2019dyp,Crichigno:2020ouj}. However, with a critical eye on universality arguments, we ask: {\it  How much of the full entropy of AdS black holes, including all corrections,  can be recovered having only knowledge of the near-horizon geometry?} Clearly, the microscopic foundations have made used of the full UV description which, via the AdS/CFT correspondence, is equivalent to the respective dual field theories. We need, however, to understand the minimal set of data that allows to understand the entropy and some of its corrections.  Given that Sen's entropy function formalism  is completely rooted in  near-horizon symmetries it becomes the ideal tool to address the question of how much of the higher-derivative corrected entropy of AdS black holes can be determined from knowledge of the near-horizon geometry alone. In this manuscript we tackle this question by focusing on higher derivative corrections to the entropy of asymptotically AdS$_4$ black holes. After a brief review of the formalism in \ref{Sec:Review}, we address the static case in Section \ref{Sec:Static}. In Section \ref{Sec:Rotating} we tackle  rotating black holes. In  Section \ref{Sec:Conclusions} we conclude that Sen's entropy function formalism seems sufficient to capture higher-derivative corrections to the entropy albeit naturally missing certain relations arising from global properties of the full black hole solution.  We also point out a number of interesting future directions.

\section{Review of  Sen's entropy function formalim}\label{Sec:Review}

In this section we briefly review Sen's prescription \cite{Sen:2005wa} to compute the entropy  of extremal black holes. In this context, extremal black holes are assumed to have an AdS$_2$ factor in the near-horizon geometry. This prescription is directly applicable to static black holes and, with some  modifications, to  rotating ones. We start reviewing with the former. \par 
To compute the entropy of a black hole in Sen's formalism, we start with an action: 
\begin{equation}
S=\int d^4 x \sqrt{-g} \mathcal{L}\left[g_{\m\n},\Phi^{(i)}, F_2^{(j)} \right]. 
\end{equation}
The Lagrangian is a functional of different fields, for example: the metric $g_{\m\n}$, scalar fields $\Phi^{(i)}$ and $1$-form gauge fields $A^{(j)}_1$ with their corresponding field strength $F^{(j)}_2 $, where $(i,j)$ run over the number of corresponding fields. We assume that the black hole is extremal and, therefore, the near-horizon geometry has $SO(1,2)\times SO(3)$ symmetry. The most general ansatz consistent with these symmetries is given by: 
\begin{align}
& ds^2=v_1\left(-r^2 dt^2+\frac{dr^2}{r^2}\right) +v_2 \left(d\t^2 +\sin^2\t d\phi^2 \right), \label{ansatzmetric}\\
& \Phi^{(i)}=u_i, \label{ansatzscalar}\\
& F_{rt}^{(j)}=-F_{tr}^{(j)}=e_{j},\quad  \ F_{\t\phi}^{(j)}=-F_{\phi \theta}^{(j)}=\frac{p_j}{4\pi} \sin\theta.  \label{ansatzF}
\end{align}
Sen's algorithm to compute the black hole entropy consists of the following three steps:
\begin{enumerate}
\item From the Ansatz \eqref{ansatzmetric}-\eqref{ansatzF}, we evaluate the Lagrangian $\mathcal{L}\left[g_{\m\n},\Phi^{(i)}, F_2^{(j)} \right]$. We then integrate over the angular coordinates and get: 
\begin{equation}
f\left( v_1,v_2,u_i,e_j,p_j \right)=\int d\t d\phi \sqrt{-g}\mathcal{L}\left[g_{\m\n},\Phi^{(i)}, F_2^{(j)} \right] . 
\end{equation} 
\item From this function we can find the entropy function which is defined as the Legendre transform of $f\left( v_1,v_2,u_i,e_j,p_j \right)$ with respect to the parameters $e_j$ labeling the electric fields. More precisely, the entropy function is
\begin{equation}
F\left(q_j, v_1,v_2,u_i,e_j,p_j  \right)=2\pi \left[ q_j e_j- f\left( v_1,v_2,u_i,e_j,p_j \right)\right]
\end{equation}
where the conjugate variable (electric charge) $q_j$ is defined as
\begin{equation}
\frac{\partial f}{\partial e_j}=q_j . \label{conjugate}
\end{equation}
\item Extremization of $F\left(q_j, v_1,v_2,u_i,e_j,p_j  \right)$, together with Eq. \eqref{conjugate} give rises to the so-called attractor equations: 
\begin{equation}
 \frac{\partial F }{\partial v_1}=\frac{\partial F }{\partial v_2}=\frac{\partial F }{\partial u_i}=\frac{\partial F }{\partial e_j}=0 . \label{attractors}
 \end{equation} 
 The number of these attractor equations is the same as the number of unknowns $(v_1, v_2, u_i, e_j )$. Therefore Eqs. \eqref{attractors} can be solved to find $(v_1, v_2, u_i, e_j )$ in terms of $q_j$ and $p_j$. The black-hole entropy is given by the value of the entropy function on these solutions: 
 \begin{equation}
 S_{Sen}= F\left(q_j, v_1,v_2,u_i,e_j,p_j  \right)\vline_{\text{solutions of Eqn. \eqref{attractors}}}. 
 \end{equation}
\end{enumerate}\par 
For rotating black holes, the near-horizon symmetry is no longer $SO(1,2)\times SO(3)$, therefore, one needs to modify the Ansatz. A systematic  discussion of the rotating  case was provided in \cite{Astefanesei:2006dd} and starts with the following Ansatz: 
\begin{align}
  & ds^2=\Omega^2(\theta) e^{2 \Psi (\theta)} \left(-r^2 dt^2+\frac{dr^2}{r^2}+\b^2 d\theta^2 \right)+e^{-2 \Psi (\theta)} \left(d\phi-\a r dt \right)^2, \\
  & \Phi_i (\theta)=u_i(\theta), \\
  & F_j=\left[e_j-\alpha  b_j(\theta \right)]dr\wedge dt + b_j'(\theta )d\theta \wedge (d\phi -\alpha r dt)
\end{align} 
where $\a$ is the rotation parameter,  a particular choice of the $\theta$ coordinate has been made and the Bianchi identity for the field strengths, $F_j$, has been implemented. To compute the black hole entropy we  follow steps similar to those performed in the case of a static black hole. The slight generalization looks as follows. One defines 
\begin{equation}
f\left[\a,\b, u_i, e_j, \Psi(\t),\Omega(\t),b_j(\t) \right]=2\pi \int d\theta \sqrt{-g} \mathcal{L}\left[g_{\m\n},\Phi_{(i)}, F_{(j)} \right]  
\end{equation}
which is a function of $(e,\a,\b, u_i, e_j)$ and a functional of $\left( \Psi(\t),\Omega(\t),b_j(\t) \right)$. Consequently, the entropy functional
\begin{equation}
F\left[q_j,J, \a,\b, u_i, e_j, \Psi(\t),\Omega(\t),b_j(\t) \right]=2\pi \left(J\a+e_j q_j - f\left[\a,\b, u_i, e_j, \Psi(\t),\Omega(\t),b_j(\t) \right] \right)
\end{equation}
is a function of $(q_j,e_j,J, \a,\b, u_i, e_j)$ and a functional of $\left( \Psi(\t),\Omega(\t),b_j(\t) \right)$. The attractor equations are: 
\begin{align}
\frac{\partial F}{\partial \a}=0, \  \frac{\partial F}{\partial \b}=0, \ \frac{\partial F}{\partial u_i}=0, \ \frac{\partial F}{\partial e_j}=0, \ \frac{\delta F}{\delta \Psi}=0, \ \frac{\delta F}{\delta \Omega}=0, \   \frac{\delta F}{\delta b}=0. \label{attractorsrot}
\end{align} 
These equations should be supplemented with appropriate boundary conditions, including those enforcing regularity of the solution. We will discuss the boundary conditions when relevant. For the rotating case, the black hole entropy is given by: 
 \begin{equation}
 S_{Sen}= F\left[q_j,J, \a,\b, u_i, e_j, \Psi(\t),\Omega(\t),b_j(\t) \right] \vline_{\text{solutions of Eq, \eqref{attractorsrot}}}. 
 \end{equation}

Sen's entropy function formalism has been very successful in capturing higher derivative corrections in asymptotically flat black holes \cite{Sen:2005iz,Sen:2007qy,Sen:2014aja}. We are interested in asymptotically AdS black holes, there are two studies discussion the application of Sen's entropy function formalism to AdS black holes \cite{Morales:2006gm,Goulart:2015lwd}. As will become clear in the next two sections, one technical contribution we provide  consists in  considering AdS$_4$ black holes with arbitrary horizon topology and with more general higher-derivative terms than those considered  previously in the literature \cite{Morales:2006gm,Goulart:2015lwd}. Our second novel contribution to the literature is to consider rotating AdS$_4$ black holes and their higher-derivative contributions. 

\section{Entropy  for static AdS black hole via Sen's formalism }\label{Sec:Static}

In this section we consider static AdS$_4$ black holes with horizon topology given by a genus  $g$ Riemann  surface. The near-horizon background symmetry implies that the metric and electromagnetic fields take the following general form
\begin{equation}
ds^2=v_1 ds^2_{AdS_2}+v_2 ds^2_{\Sigma_g} , \quad \ F_{tr}=-F_{rt}=e
\end{equation}
where $ds^2_{\Sigma_g} $ is the metric on the genus $g$  Riemann surface, $\Sigma_g$.

One of our main goals is to incorporate higher-curvature corrections to the entropy. We naturally start by recalling a number of  relevant formulas from differential geometry. For any two-dimensional surface the Riemann tensor and Ricci scalar  satisfy:  
\begin{align}
R_{\m\n\r\s}=\frac{R}{2} \left(g_{\m\r}g_{\n\s}-g_{\m\s}g_{\n\r} \right), \ R=2 K \ . 
\end{align}
We recall that the Gauss-Bonnet term is defined as 
\begin{equation}
GB=R_{\m\n\r\s}R^{\m\n\r\s}-4 R_{\m\n}R^{\m\n}+R^2.
\ee
For the particular case at hand, the metric is block-diagonal \footnote{By capital Latin letters we will denote the full 4d coordinates. Small Latin letters will denote the AdS$_2$ coordinates $(t,r)$ and Greek letters will denote the coordinates of $\Sigma_g$: $(\t,\phi)$. }. So we can write the following
\begin{align}
& g_{AB}=g_{ab}\oplus g_{\m\n}, \\
& R_{AB}=R_{ab}\oplus R_{\m\n},\\
&R_{ABCD}=R_{abce}\oplus R_{\m\n\r\s}, \\
& R=R_{AdS_2}+R_{\Sigma_g}=-\frac{2}{v_1}+\frac{R^{unit}_{\Sigma_g} }{v_2}, 
\end{align}
where $R^{unit}_{\Sigma_g}$ is the scalar curvature of the Riemann surface $\Sigma_g$ with unit radius. Direct evaluation yields
\begin{align}
GB&=R_{ABCD} R^{ABCD}-4 R_{AB}R^{AB}+R^2 \nn\\
& =GB_{AdS_2}+GB_{\Sigma_g}+2 R_{AdS_2} R_{\Sigma_g} \nn \\
& =2 R_{AdS_2} R_{\Sigma_g}=-\frac{4}{v_1 v_2} R^{unit}_{\Sigma_g},  
\end{align}
where in the third line we have used the fact that for any two-dimensional manifold, the Gauss-Bonnet term is identically zero due to the relation between the Riemann and Ricci tensors with the metric.

Having cleared the geometric preliminaries, we return to Sen's formalism whose starting point is the action, 
\begin{align}
S=\frac{1}{16\pi G_4}\int d^4x \sqrt{-g} \left(R+\Lambda-\frac{1}{4}F^2+a GB \right).
\end{align}
We calculate the function
\begin{align}
f(v_1,v_2,e)&=\int d\theta d\phi \sqrt{-g} \mathcal{L}\nn \\
&=\frac{v_1v_2}{16\pi G_4} \left\{ S\left(\Sigma_g^{unit} \right) \left(-\frac{2}{v_1}+\Lambda+\frac{e^2}{2 v_1^2} \right)+\left(\frac{1}{v_2}-\frac{4a}{v_1 v_2} \right)4\pi \chi\left(\Sigma_g \right) \right\}
\end{align}
where $S\left(\Sigma_g^{unit} \right) $ and $\chi\left(\Sigma_g \right) $ are the surface area and Euler characteristic of the Riemann surface $\Sigma_g^{unit} $. In obtaining the above equation we have used the Gauss-Bonnet theorem for a Riemann surface: 
\begin{equation}
\int d\theta d\phi \sqrt{g_{\Sigma_g}} R_{\Sigma_g}=4\pi \chi\left(\Sigma_g \right). 
\end{equation}
\par 
Now we can construct the entropy function  \`a la Sen 
\begin{align}
F(q,v_1,v_2,e)&=2\pi \left[eq-f(v_1,v_2,e) \right]\nn \\
&=2\pi \left[eq- \frac{v_1v_2}{16\pi G_4} \left\{ S\left(\Sigma_g^{unit} \right) \left(-\frac{2}{v_1}+\Lambda+\frac{e^2}{2 v_1^2} \right)+\left(\frac{1}{v_2}-\frac{4a}{v_1 v_2} \right)4\pi \chi\left(\Sigma_g \right) \right\} \right].
\end{align}
This function should be extremized with respect to $v_1,v_2,e$:
\begin{equation}
\frac{\partial F}{\partial v_1}=\frac{\partial F}{\partial v_2}=\frac{\partial F}{\partial e}=0 \ . 
\end{equation}
These give the following equations: 
\begin{eqnarray}
 S\left(\Sigma_g^{unit} \right) e^2  v_2-2 v_1^2 \left\{\Lambda  S v_2+4 \pi  \chi\left( \Sigma_g^{unit}\right) \right\}&=&0, \\
  e^2+2 v_1 (\Lambda  v_1-2)&=&0, \\ 
 2 \pi  q-\frac{ S\left(\Sigma_g^{unit} \right) e v_2 }{8 G_4 v_1 }&=&0. 
\end{eqnarray}
The solutions to these equations are given by: 
\begin{align}
v_1=\frac{S v_2}{2\pi\chi +Sv_2 \Lambda}, \ e=\frac{\sqrt{2 S v_2 \left(4\pi\chi+S v_2 \Lambda \right)}}{2\pi\chi+S v_2 \Lambda}, \ q=\frac{\sqrt{ S v_2 \left(4\pi\chi+S v_2 \Lambda \right)}}{8\sqrt{2} G_4 \pi}. 
\end{align} 
Inserting these values into the function $F(q,v_1,v_2,e)$ we find 
\begin{align}
S_{BH}&=\frac{Sv_2}{4 G_4}+\frac{2\pi a\chi}{G_4} \nn \\
&= \frac{A(H)}{4 G_4}+\frac{2\pi a\chi (H)}{G_4} . \label{Bhcorrected}
\end{align}
We can observe that the Bekenstein-Hawking formula gets corrected in the presence of a Guss-Bonnet type higher-derivative correction which is tracked by the parameter $a$ introduced in the action.

\subsection{Agreement with higher curvature corrections in supergravity}

In this subsection we  compare the results obtained using Sen's entropy function formalism with a recent study which discussed the influence of various higher-derivative terms, motivated by conformal supergravity, on the  the Wald entropy \cite{Bobev:2020egg} for certain black holes in AdS$_4$. Let us briefly summarize  the  main result of \cite{Bobev:2020egg}. The  starting point  of \cite{Bobev:2020egg} is the following action which includes higher-derivative terms as follows:
\begin{equation}\label{Eq:HD-Action}
\mathcal{L}=\mathcal{L}_{2\partial}+(c_1-c_2) \mathcal{L}_{W^2}+c_2 \mathcal{L}_{GB},
\end{equation}
where 
\begin{align}
& \frac{1}{\sqrt{g}} \mathcal{L}_{2\partial}=-\frac{1}{16\pi G_N} \left[ R+\frac{6}{\ell^2}-\frac{1}{4} F_{\m\n} F^{\m\n}\right] \label{2deri}\\
& \frac{1}{\sqrt{g}} \mathcal{L}_{W^2}=\left( C_{\m\n}^{\ \ \r\s}\right)^2  -\frac{F_{\m\n} F^{\m\n}}{\ell^2} +\frac{1}{2} \left( F_{\m\n}^{+}\right)^2 \left( F_{\r\s}^{-}\right)^2 -4 F_{\m\n}^{-} R^{\m\r} F_{\r}^{+\ \n} \nn \\
&\qquad\qquad\qquad +8 \left( \nabla^\m F_{\m\n}^{-} \right) \left( \nabla^\r F_{\r}^{+\n} \right), \label{W2} \\
& \frac{1}{\sqrt{g}} \mathcal{L}_{GB}=R_{\m\n\r\s}R^{\m\n\r\s}-4 R_{\m\n}R^{\m\n}+R^2 \ .\label{GB}
\end{align}
In the above expressions $F^\pm$ are the self-dual and anti self-dual parts of the field-strength tensor, defined as: 
\begin{equation}
F^\pm_{\m\n}=\frac{1}{2}\left( F_{\m\n}\pm \frac{1}{2} \epsilon_{\m\n\r\s} F^{\r\s} \right)\ . 
\end{equation}
We recall some identities that allow to simplify the higher-derivative action presented in Eq. (\ref{Eq:HD-Action}). In $d$-dimensions, the Weyl tensor is defined as: 
\begin{align}
C_{\m\n\r\s }& =R_{\m\n\r\s}-\frac{2}{d-2} \left( g_{\r[\m} R_{\n]\s}-g_{\s[\m}R_{\n]\r} \right)+\frac{2}{(d-1)(d-2)} R g_{\r[\m}g_{\n]\s}. \nn
\end{align}
From this we can find the Weyl squared as
\begin{align}
C_{\m\n\r\s} C^{\m\n\r\s}&=R_{\m\n\r\s} R^{\m\n\r\s}-\frac{4}{d-2} R_{\m\n} R^{\m\n}+\frac{2}{(d-1)(d-2)}R^2 \nn \\
& \overset{d=4}{=}R_{\m\n\r\s} R^{\m\n\r\s}-2 R_{\m\n} R^{\m\n}+\frac{1}{3} R^2 \nn \\
& = GB+2 R_{\m\n}R^{\m\n}-\frac{2}{3}R^2 \ .  \label{Weyl2}
\end{align}
The two equations of motion from the $2$-derivative Lagrangian \eqref{2deri} are:
\begin{align}
& R_{\m\n}-\frac{1}{2} R g_{\m\n} -\frac{3}{\ell^2} g_{\m\n}+\frac{1}{8} g_{\m\n} F_{\a\b}F^{\a\b}-\frac{1}{2} F_{\m\a}F_{\n}^{\ \a}=0 \label{EOM1}, \\
& \nabla_\mu F^{\m\n}=0 \label{EOM2} . 
\end{align}
Contracting Eq. \eqref{EOM1} with $g^{\m\n}$ we find 
\begin{equation}
R=-\frac{12}{\ell^2} \ ,
\end{equation}
implying that the metric has constant Ricci curvature. Inserting back this into Eq. \eqref{EOM1} we can write the Ricci tensor
\begin{equation}
R_{\m\n}=-\left( \frac{3}{\ell^2}+\frac{1}{8} F_{\a\b}F^{\a\b} \right)g_{\m\n}+\frac{1}{2}F_{\m\a} F_{\n}^{\ \a} \ . 
\end{equation}
Using the value of Ricci curvature and from Eq. \eqref{2deri} we can write
\begin{equation}
\frac{F_{\m\n} F^{\m\n}}{\ell^2}=\frac{64\pi G_N}{\sqrt{g} \ell^2} \mathcal{L}_{2\partial}-\frac{24}{\ell^4} \ . \label{F2}
\end{equation}
Inserting Eqs. \eqref{Weyl2} and \eqref{F2} into Eq. \eqref{W2} we find: 
\begin{align}
\frac{1}{\sqrt{g}}\mathcal{L}_{W^2}&=\frac{1}{\sqrt{g}} \mathcal{L}_{GB}-\frac{64\pi G_N}{\sqrt{g} \ell^2} \mathcal{L}_{2\partial}+2 R_{\m\n}R^{\m\n} -\frac{72}{\ell^4}+ \frac{1}{2} \left( F_{\m\n}^{+}\right)^2 \left( F_{\r\s}^{-}\right)^2 -4 F_{\m\n}^{-} R^{\m\r} F_{\r}^{+\ \n} \nn \\
&\qquad\qquad\qquad +8 \left( \nabla^\m F_{\m\n}^{-} \right) \left( \nabla^\r F_{\r}^{+\n} \right).  
\end{align} 
When the fields satisfy the equations of motion, the following relation is true \cite{Bobev:2020egg}:
\begin{equation}
I_{W^2}=I_{GB}-\frac{64\pi G_N}{\ell^2} I_{2\partial} \ . 
\end{equation}
where $I$ denotes the on-shell quantity. In this case the higher derivative Lagrangian 
\begin{equation}
{\cal L}_{HD}={\cal L}_{2\partial}+(c_1-c_2) {\cal L}_{W^2}+c_2 {\cal L}_{GB}
\end{equation}
on-shell reduces to 
\begin{equation}
I_{HD}=\left(1+\a \right) I_{2\partial}+c_1 I_{GB}
\end{equation}
where 
\begin{equation}
\a=\frac{64\pi G_N}{L^2} \left( c_2-c_1 \right) \ . 
\end{equation}
The Wald entropy in this case has been computed and the formula is given in Eq. (14) of \cite{Bobev:2020egg} which we reproduce here
\begin{equation}
S_{BH(Wald)}=(1+\a) \frac{A_H}{4 G_N}-32 \pi^2 c_1 \chi (H) \label{SBHBobev}
\end{equation}
where $A_H$ and $\chi (H)$ are the area and Euler character of the horizon, respectively.

To compare our result in the previous section with the one of Bobev et al \cite{Bobev:2020egg} we need to make the following mapping: 
\begin{equation}
G_4=-\frac{G_N}{1+\a}, \ \ a=-\frac{16\pi G_N}{1+\a} c_1 \ . 
\end{equation}
which gives 
\begin{equation}
S_{BH}=-\left(1+\a \right) \frac{A(H)}{4 G_N}+32\pi^2 c_1 \chi (H)
\end{equation}
which is exactly Eq. (14) of \cite{Bobev:2020egg} up to an overall minus sign. The reason for the overall minus sign is because the definition of Wald's entropy used in \cite{Bobev:2020egg} has a relative minus sign with respect to  the definition used by Sen in \cite{Sen:2005wa}. Therefore, we have correctly reproduced Wald's entropy following Sen's entropy function formalism. 

In the context of the AdS/CFT correspondence one often thinks of the geometry as realizing an RG flow in the dual field theory setup. Namely, the near-horizon region roughly corresponds to the IR theory while the  higher dimensional AdS$_4$ asymptotic region corresponds to the UV side of the theory. The computation we have performed demonstrates that for these type of configurations, the near-horizon geometry (IR data) is enough to reproduce the entropy, including higher-derivative corrections. Most field theory computations determine the entropy starting from an observable which is, {\it a priori}, defined in the UV theory  \cite{Benini:2015eyy,Choi:2019miv,Choi:2019zpz, Kantor:2019lfo, Nahmgoong:2019hko, Nian:2019pxj,Bobev:2019zmz,Benini:2019dyp,Crichigno:2020ouj}. Therefore, the effectiveness of Sen's entropy function formalism for AdS$_4$ black holes indicates that in the field theory side there should be an IR sector from which the entropy can be computed. Moreover, Sen's entropy function formalism for AdS has the added technical advantage of being able to reproduce the entropy, including all higher-derivative corrections, with knowledge of the near-horizon geometry alone.

\section{Rotating AdS$_4$ black holes in the entropy function formalism}\label{Sec:Rotating}

In this section we consider rotating AdS$_4$ black holes. Such discussion is lacking in the literature of Sen's entropy function formalism for AdS spacetimes. In particular, Morales and Samtleben  in \cite{Morales:2006gm} as well as  Goulart in  \cite{Goulart:2015lwd} considered only static AdS$_4$ black holes\footnote{Morales and Samtleben did consider rotating AdS$_5$ black holes but that problem is in different symmetry class than rotating AdS$_4$ black holes}. There are, however, recent discussions of the entropy function for rotating asymptotically AdS$_4$ black holes, see for example, \cite{Choi:2018fdc,Cassani:2019mms}. These works evaluated the entropy function using the full solutions obtained  originally in  \cite{Chong:2004na} and further discussed in \cite{Cvetic:2005zi,Chow:2013gba, Choi:2018fdc,Cassani:2019mms}.  We will compare the results of Sen's AdS entropy function formalism with the direct result arising from the full solution later in this section.

In order to clarify various notions specific to rotating, electrically charged,  asymptotically AdS$_4$ black holes, we  will first briefly review a number of simpler cases. We will first review rotating, electrically charged black holes  with a Gauss-Bonnet term in flat spacetime  \ref{SubSec:FlatRotGB} before turning to the main result for the rotating, electrically charged asymptotically AdS$_4$ black holes in \ref{SubSec:AdSRotGB}. We will conclude this section showing in \ref{SubSec:SugraAdS4} that the solution to the near-horizon geometry obtained \`a la Sen matches precisely with the Bardeen-Horowitz limit of known, rotating,  electrically charged,  BPS black holes.

\subsection{Rotating Black Holes in Flat Space}\label{SubSec:FlatRotGB}

As reviewed in section \ref{Sec:Review}, the standard formulation of Sen's entropy function formalism utilizes the symmetries of the AdS$_2$ near-horizon geometry. For the case of rotating black holes the situation requires important modifications first described in \cite{Astefanesei:2006dd} and summarized in section \ref{Sec:Review}. 

Let us start by considering the calculation of the entropy of rotating black holes in asymptotically flat four-dimensional spacetime. To capture the effect of higher-curvature corrections, we include a  Gauss-Bonnet term in the action. It is well known that in four dimensions, the Gauss-Bonnet term is purely topological and therefore, does not contribute to the equations of motion.  Our starting point is the action:  
\begin{align}
S&=\frac{1}{16\pi G_4}\int d^4x \sqrt{-g} \left(R+a GB \right), \\
& =\int d^4x \sqrt{-g}  \mathcal{L}\left[g_{\m\n}\right],
\end{align}
we will use Sen's formalism as modified in \cite{Astefanesei:2006dd}. We take the following metric Ansatz: 
\begin{align}
ds^2=\Omega^2(\theta) e^{2 \Psi (\theta)} \left(-r^2 dt^2+\frac{dr^2}{r^2}+\b^2 d\theta^2 \right)+e^{-2 \psi (\theta)} \left(d\phi-\a r dt \right)^2 \ . \label{metricrotating}
\end{align}
In  Sen's formalism we first compute  $\mathcal{L}$ for the metric Ansatz above, then we construct: 
\begin{equation}
f\left[ \a,\b,\Psi(\theta), \Omega (\theta) \right] =2\pi \int_0^\pi d\theta \sqrt{-g} \mathcal{L}\left[ \a,\b,\Psi(\theta), \Omega (\theta) \right]. 
\end{equation}
The entropy functional is defined as: 
\begin{equation}
F\left[J,\a, \b,\Psi(\theta), \Omega (\theta) \right]=2\pi \left( J\a-f\left[ \a,\b,\Psi(\theta), \Omega (\theta) \right]  \right) . 
\end{equation}
This functional should be extremized  according to: 
\begin{align}
\frac{\partial F}{\partial \a}=0, \  \frac{\partial F}{\partial \b}=0, \ \frac{\delta F}{\delta \Psi}=0, \ \frac{\delta F}{\delta \Omega}=0. \label{extremeeq}
\end{align} 
This extremization, in general, will give differential equations for $\Psi(\theta)$ and $\Omega(\theta)$. To solve these uniquely we need to impose boundary conditions at $\theta=0, \pi$. The boundary conditions we impose  follow \cite{Astefanesei:2006dd} and imply regularity of the background:
\begin{align}
\Omega(\theta) e^{\Psi(\theta)} \to \text{constant as} \ \theta\to 0,\pi \\
\b \Omega(\theta) e^{2 \Psi(\theta)} \sin\theta \to 1 \ \text{as} \ \theta\to 0,\pi. 
\end{align}
\par 
 From the metric ansatz \eqref{metricrotating} we can obtain
\begin{align}
\sqrt{-g} R& = \frac{\alpha ^2 \beta ^2 e^{-4 \Psi (\theta )}-4 \Omega (\theta )^2 \left(\beta ^2+\Psi '(\theta )^2\right)+4 \Omega '(\theta )^2}{2 \beta  \Omega (\theta )}-2 \frac{\partial }{\partial \theta } \left[ \frac{\Omega (\theta ) \Psi '(\theta )+2 \Omega '(\theta )} {\beta } \right] 
\end{align}
The Gauss-Bonnet term can be written as: 
\begin{align}
\sqrt{-g} GB=\frac{d}{d\theta}\left[  \frac{2 e^{-6 \Psi (\theta )}}{\beta ^3 \Omega (\theta )^4} \right.  & \left[ \alpha ^2 \beta ^2 \left\{ 5 \Omega (\theta ) \Psi '(\theta )+2 \Omega '(\theta )\right\} \right.  -4 e^{4 \Psi (\theta )} \Omega (\theta ) \Psi '(\theta ) \left\{ \Omega (\theta )^2 \left(\beta ^2+\Psi '(\theta )^2\right) \right. \nn \\
& +2 \Omega (\theta ) \Psi '(\theta ) \Omega '(\theta )+\Omega '(\theta )^2 \left. \right\} \left. \right] \left. \frac{}{} \right] , 
\end{align} 
and is a total derivative, as expected.  Therefore, we can write 
\begin{align}
f\left[\a,\b,\Psi(\t)\Omega(\t) \right]=\frac{1}{8 G_4} \int d\theta  \frac{\alpha ^2 \beta ^2 e^{-4 \Psi (\theta )}-4 \Omega (\theta )^2 \left(\beta ^2+\Psi '(\theta )^2\right)+4 \Omega '(\theta )^2}{2 \beta  \Omega (\theta )}+BT
\end{align}
where $BT=BT_{EH}+BT_{GB}$ are the boundary terms. In particular, they are: 
\begin{align}
& BT_{EH}=-\frac{1}{4 G_4} \left[ \Omega(\theta) e^{2\Psi(\theta)} \sin\t \left\{ \Omega (\theta ) \Psi '(\theta )+2 \Omega '(\theta ) \right\} \right]_0^\pi \\
& BT_{GB}=\frac{a}{4 G_4} \left[ \sin\theta\left\{ -4 e^{4 \Psi (\theta )} \sin ^2(\theta ) \Psi '(\theta ) \left(\Omega (\theta ) \Psi '(\theta )+\Omega '(\theta )\right)^2-4 \Psi '(\theta )  \right.  \right. \nn \\
& \qquad\qquad\qquad\qquad\qquad\qquad\qquad\left. \left. +\frac{\alpha ^2 e^{-4 \Psi (\theta )} \left(5 \Omega (\theta ) \Psi '(\theta )+2 \Omega '(\theta )\right)}{\Omega (\theta )^3}\right\} \right]_0^\pi .
\end{align}
Since the Gauss-Bonnet term is a total derivative, it does not change the equations of motion. In particular, solutions of the $2$-derivative equations of motion obtained from the Einstein-Hilbert action will be the full solutions of the system. The solutions  to  the extremization problem with the appropriate boundary conditions were obtained in \cite{Astefanesei:2006dd} in units where  $\frac{1}{16\pi G_4}=1$ and we record it below: 
\begin{align}
\a=1,  \quad  \Omega(\t)= \frac{J}{8\pi} \sin\t,  \quad \ e^{-2 \Psi (\t)}=\frac{J}{4\pi} \frac{\sin^2\t}{1+\cos^2\t}.
\end{align}
In this case the black hole entropy is: 
\begin{equation}
S_{Sen}= 2\pi J+64\pi^2 a \ . 
\end{equation}
The first term is is the usual Bekenstein-Hawking entropy formula. The second term is our main contribution to this discussion, it arises due to the Gauss-Bonnet and it can be written as: 
\begin{equation}
 64\pi^2 a=32\pi^2 a \chi \left(H \right). 
 \end{equation} 
 This matches the formula  \eqref{Bhcorrected} and reinforces its universality. 

 \subsubsection*{Including a gauge field}
 Let us finish this subsection by including a gauge field in the computation just sketched. We now consider the Einstein-Maxwell theory together with the Gauss-Bonnet term whose action  is: 
 \begin{equation}
 S=\int d^4 x\sqrt{-g}\left(R-\frac{1}{4}F_{\m\n}F^{\m\n}+a GB \right). 
 \end{equation}
 We use the Ansatz used in \cite{Astefanesei:2006dd} meaning that the metric remains the same and, therefore, all the geometrical formulas for  the Gauss-Bonnet term are unchanged. The appropriate solution to the coupled system of differential equations  are again given in \cite{Astefanesei:2006dd} and we record them here: 
 \begin{align}
&  \b=1, \quad  \alpha =\frac{J}{\sqrt{J^2+\left(\frac{q^2}{8 \pi }\right)^2}}, \quad  \Omega(\t)=\frac{\sqrt{J^2+\left(\frac{q^2}{8 \pi }\right)^2}}{8 \pi } \sin\t, \nn \\
& \Psi (\theta)=-\frac{1}{2} \log \left(\frac{2 a \sin ^2(\theta )}{\cos ^2(\theta )+\frac{q^2 \sin ^2(\theta )}{8 \pi  \sqrt{J^2+\left(\frac{q^2}{8 \pi }\right)^2}}+1}\right) . 
 \end{align}
 Following the same procedure as before we compute the entropy: 
 \begin{equation}
 S_{Sen}=2\pi \sqrt{J^2+\left(\frac{q^2}{8 \pi }\right)^2} +64\pi^2 a. 
 \end{equation}
 As before the first term is the usual Bekenstein-Hawking term and the second one is the contribution from Gauss-Bonnet. Again, this matches with Eq.  \eqref{Bhcorrected}.

 \subsection{Kerr-AdS black holes}\label{SubSec:AdSRotGB}
  In this section we study the  Kerr-AdS black hole and in particular we  examine how the entropy of the Kerr-AdS black hole can be reproduced from Sen's entropy function formalism.   When the parameters of the black hole satisfy certain conditions, the near-horizon geometry contains an AdS$_2$ factor. We will operationally refer to the presence of such AdS$_2$ near-horizon subspace  as`` extremal  \`a la Sen''.   \par
  Our starting point is the following ansatz: 
  \begin{equation}
   ds^2=v_1(\theta)\left[-(1+r^2)dt^2+\frac{dr^2}{1+r^2} \right]+\b (\theta)d\theta^2+\frac{\lambda \sin^2\theta}{\b(\theta)} \left(d\phi-\alpha r dt \right)^2 \label{metricKerr-AdS}
   \end{equation} 
   where $v_1 (\theta), \b (\theta)$ are metric functions and $\l, \a$ are constants. The metric on the $(\theta,\phi)$ plane is: 
   \begin{align}
   ds^2_{\theta-\phi}&=\b(\theta)d\theta^2+\frac{\lambda \sin^2\theta}{\b(\theta)}d\phi^2, \\
   &= \beta(\theta)\left[ d\theta^2+\frac{\lambda \sin^2\theta}{\b^2(\theta)} d\phi^2 \right]. 
   \end{align}
   We demand this to be a smooth variation of the metric of $S^2$. It requires: 
   \begin{equation}
   \lim_{\theta\to 0,\pi} \frac{\l}{\b^2(\t)}=1 . \label{lambdacondition}
   \end{equation}
   These conditions should determine the value of $\l$. This also requires that 
   \begin{equation}
   \lim_{\t\to 0} \b(\theta)=\lim_{\t\to \pi} \b(\theta). 
   \end{equation}
   \par 
   The action we consider is: 
   \begin{equation}
   S=\int d^4x \sqrt{-g} \left[R+\frac{6}{\ell^2}+\gamma GB \right]. 
   \end{equation}
   Now we compute: 
   \begin{align}
   f\left[\a, v_1(\theta),\b(\theta) \right]&=\int d\phi d\theta \sqrt{-g} \left[R+\frac{6}{\ell^2}+\gamma GB \right],\nn \\
  & =2\pi \int d\theta \mathcal{L}\left[\a, v_1(\theta),\beta(\theta) \right]+2\pi BT.  
   \end{align}
   The Lagrangian is given by: 
   \begin{align}
  \mathcal{L}\left[\a, v_1(\theta),\beta(\theta) \right]&=  -2 \sqrt{\b(\t)v_2 (\t)}+\frac{\a^2}{2}  \sqrt{\b(\t)v_2 (\t)} \frac{v_2(\t)}{v_1(\t)}+\frac{\sqrt{v_2(\theta )} v_1'(\theta )^2}{2 \sqrt{\beta (\theta )} v_1(\theta )} \nn \\
  & \ \ +\frac{v_1'(\theta ) v_2'(\theta )}{\sqrt{\beta (\theta )} \sqrt{v_2(\theta )}}  +\frac{6}{\ell^2} \sqrt{\lambda } \sin (\theta ) v_1(\theta ) \label{Lag}
   \end{align}
   where we have defined 
   \begin{equation}
   v_2(\t)=\frac{\l \sin^2\t}{\b(\t)}
   \end{equation}
   for brevity. The boundary terms $BT=BT_{EH}+BT_{GB}$ arise due to the total derivates. There are two kinds of boundary term, one from the Einstein-Hilbert term of the action and the other one from  the GB term. The boundary terms are: 
   \begin{align}
&  BT_{EH}=-\left[\frac{v_1(\theta ) v_2'(\theta )}{\sqrt{\beta (\theta )} \sqrt{v_2(\theta )}}+\frac{2 \sqrt{v_2(\theta )} v_1'(\theta )}{\sqrt{\beta (\theta )}}  \right]_{\t=0}^{\t=\pi}  \\
 & BT_{GB}=-\g \left[ \frac{v_2(\theta ) \left(-v_1(\theta ) v_2'(\theta ) \left(v_1'(\theta )^2+4 \beta (\theta ) v_1(\theta )-3 \alpha ^2 \beta (\theta ) v_2(\theta )\right)-2 \alpha ^2 \beta (\theta ) v_2(\theta )^2 v_1'(\theta )\right)}{v_1(\theta )^2 (\beta (\theta ) v_2(\theta ))^{3/2}} \right]_0^\pi.
    \end{align}
    The equations for determining $v_1(\theta)$ and $\b(\t)$ come from the extremization of the Lagrangian \eqref{Lag}. They are: 
    \begin{align}
  &   \ \frac{-\alpha ^2 \lambda  \sin ^3(\theta )+\sin (\theta ) \left(v_1'(\theta )^2-2 v_1(\theta ) v_1''(\theta )+4 v_1(\theta )^2\right)-2 \cos (\theta ) v_1(\theta ) v_1'(\theta )}{\beta (\theta ) v_1(\theta )^2}  \nn \\
&  +\frac{12 \sin (\theta )}{\ell^2}-\frac{4 \sin (\theta ) \beta '(\theta )^2}{\beta (\theta )^3} +\frac{2 \sin (\theta ) \beta ''(\theta )+\beta '(\theta ) \left(6 \cos (\theta )+\frac{2 \sin (\theta ) v_1'(\theta )}{v_1(\theta )}\right)}{\beta (\theta )^2}=0,  \label{Eq1:Kerr-AdS}  \\
&  \alpha ^2 \lambda  \sin ^3(\theta )+\sin (\theta ) \left(v_1'(\theta )^2-2 v_1(\theta ) v_1''(\theta )\right)+2 \cos (\theta ) v_1(\theta ) v_1'(\theta )=0. \label{Eq2:Kerr-AdS}
    \end{align}
    To solve these equations we use the following ansatze:
    \begin{align}
    \label{Eq:betaKerr-AdS}
    & v_1 (\t)=\frac{1}{V} \left(r_+^2 +a^2 \cos^2\t  \right), \\
    & \b (\t)= \frac{r_+^2 +a^2 \cos^2\t }{1-\frac{a^2}{\ell^2}  \cos^2\t }, 
\end{align}  
where $r_+, a, V$ are constants. Inserting these into    Eqs. \eqref{Eq1:Kerr-AdS}-\eqref{Eq2:Kerr-AdS} we find two relations between them: 
\begin{align}
& \alpha ^2 \lambda -\frac{4 a^2 r_+^2}{V^2}=0, \label{Eq3:Kerr-AdS} \\
& r_+^2 \left(a^2+\ell^2\right)-a^2 \ell^2+3 r_+^4=0. \label{Eq4:Kerr-AdS}
\end{align}
Eq. \eqref{Eq4:Kerr-AdS} is indeed the extremality condition mentioned in \cite{Dias:2012pp}. From condition \eqref{lambdacondition} we find: 
\begin{equation}
\lambda=\frac{\left(r_+^2+a^2 \right)^2}{\left(1-\frac{a^2}{\ell^2} \right)^2}. 
\end{equation}
Now we can find the entropy function a la Sen: 
\begin{equation}
F\left[\a, v_1(\t),\b (\t) \right]= 2\pi \left( J\alpha-  f\left[\a, v_1(\theta),\b(\theta) \right]  \right).\label{EFKerr-AdS}
\end{equation}
From the condition $\frac{\partial F}{\partial \alpha}=0$, we can find the angular momentum: 
\begin{equation}
J=\frac{2 \pi  \alpha  \lambda ^{3/2} V \left(\ell^2+3 r_+^2\right)}{a^2 \ell^2 r_+^2}. 
\end{equation}
Inserting this into the entropy function Eq. \eqref{EFKerr-AdS} we find 
\begin{equation}
S_{Sen}=16\pi^2 \sqrt{\lambda}+ 64\pi^2 \g . \label{SBHKerr-AdS}
\end{equation}
The first term in the above equation is the Bekenstein-Hawking part: 
\begin{equation}
S_{BH}=\frac{A_H}{4 G_N}=4\pi A_H=4\pi \sqrt{\l}\int_0^{2\pi} d\phi\int_0^\pi d\theta\sin\t=16\pi^2\sqrt{\l}. 
\end{equation}
Recall that we are using the unit where $\frac{1}{16\pi G_N}=1$, restoring Newton's constant we recover the more familiar expression $S_{BH}=\frac{\pi}{G}\frac{r_+^2+a^2}{1-a^2 \ell^{-2}}$ The second part comes from the correction due to higher derivative term: 
\begin{equation}
S_{GB}=\frac{2\pi \g \chi (H)}{G_N}=32\pi^2 \g \chi (H)
\end{equation}
where $\chi(H)$ is the Euler characteristic of the horizon. It can be obtained from Gauss-Bonnet theorem: 
\begin{equation}
4\pi \chi(H)=\int d\t d\phi \sqrt{g_H} R_H
\end{equation}
where $g_H$ and $R_H$ are the determinant and Ricci scalar of the metric of the horizon respectively. A straightforward calculation yields: 
\begin{equation}
\chi(H)=2. 
\end{equation}
So the correction to the entropy from the Gauss-Bonnet term is: 
\begin{equation}
S_{GB}= 64\pi^2 \g. 
\end{equation}
The above result reinforces the universality  of our general formula \eqref{Bhcorrected}.

\subsection{Kerr-Newman-AdS black holes} \label{KNAdS}
In this section we  finally consider the  electrically charged rotating black hole in an asymptotically AdS$_4$  spacetime.  We consider the Einstein-Maxwell-Gauss-Bonnet theory theory with a negative cosmological constant 
\begin{equation}
S=\int d^4x\sqrt{-g} \left(R+\frac{6}{\ell^2}-\frac{1}{4}F_{\m\n}F^{\m\n} +\gamma GB \right).
\end{equation}
As prescribed by  Sen's classical entropy function formalism, we by taking the following Ansatz: 
\begin{align}
& ds^2=v_1(\theta)\left(-r^2 dt^2+\frac{dr^2}{r^2} \right)+\b (\t)d\theta^2+\frac{\l\sin^2\t}{\b (\t)} \left(d\phi+\a r dt \right)^2\\
& F=-e\a b(\t)dt\wedge dr+e b'(\t)d\t\wedge (d\phi+\a r dt). 
\end{align}
This Ansatz is consistent with the near horizon geometry and satisfies the Bianchi identity $dF=0$. The metric is the same as in the previous section implying that the boundary terms are precisely those of the previous section. The  Lagrangian, however, changes to: 
  \begin{align}
 &  \mathcal{L}\left[\a, v_1(\theta),\beta(\theta) \right]=  -2 \sqrt{\b(\t)v_2 (\t)}+\frac{\a^2}{2}  \sqrt{\b(\t)v_2 (\t)} \frac{v_2(\t)}{v_1(\t)}+\frac{\sqrt{v_2(\theta )} v_1'(\theta )^2}{2 \sqrt{\beta (\theta )} v_1(\theta )} \nn \\
  & \ \ +\frac{v_1'(\theta ) v_2'(\theta )}{\sqrt{\beta (\theta )} \sqrt{v_2(\theta )}}  +\frac{6}{\ell^2} \sqrt{\lambda } \sin (\theta ) v_1(\theta )-\frac{1}{2} e^2 \sqrt{\lambda } \sin (\theta ) v_1(\theta ) \left(\frac{\csc ^2(\theta ) b'(\theta )^2}{\lambda }-\frac{\alpha ^2 b(\theta )^2}{v_1(\theta )^2}\right). \label{LagKNAdS}
   \end{align}
   From this, we can find the equations for $v_1(\t), \b(\t),$ and $b(\t)$ by taking the appropriate variation. They are given by: 
   \begin{align}
 &   \ 2 \beta (\theta )^3 \left[ e^2 \ell^2 v_1(\theta )^2 b'(\theta )^2+\lambda  \sin ^2(\theta ) \left(\alpha ^2 e^2 \ell^2 b(\theta )^2-12 v_1(\theta )^2\right)\right]+4 \lambda  \ell^2 \sin ^2(\theta ) v_1(\theta )^2 \beta '(\theta )^2\nn \\
& +\lambda  \ell^2 \beta (\theta )^2 \sin (\theta ) \left[ \alpha ^2 \lambda  \sin ^3(\theta )-\sin (\theta ) \left(v_1'(\theta )^2-2 v_1(\theta ) v_1''(\theta )+4 v_1(\theta )^2\right)+2 \cos (\theta ) v_1(\theta ) v_1'(\theta )\right]    \nn \\
& +2 \lambda  \ell^2 \beta (\theta ) \sin (\theta ) v_1(\theta ) \left[ \beta '(\theta ) \left(\sin (\theta ) v_1'(\theta )+3 \cos (\theta ) v_1(\theta )\right)+\sin (\theta ) v_1(\theta ) \beta ''(\theta )\right]=0, \label{Eq1KNAdS}\\
&  \ \alpha ^2 \lambda  \sin ^3(\theta )+\sin (\theta ) \left(v_1'(\theta )^2-2 v_1(\theta ) v_1''(\theta )\right)+2 \cos (\theta ) v_1(\theta ) v_1'(\theta )=0, \label{Eq2KNAdS}\\
&  \ \csc ^2(\theta ) v_1(\theta ) \left[ b'(\theta ) \left(v_1'(\theta )-\cot (\theta ) v_1(\theta )\right)+v_1(\theta ) b''(\theta )\right]+\alpha ^2 \lambda  b(\theta )=0. 
   \end{align}
   Solutions of these equations are: 
   \begin{align}
  & v_1 (\t)=\frac{1}{V} \left(r_+^2 +a^2 \cos^2\t  \right), \\
    & \b (\t)= \frac{r_+^2 +a^2 \cos^2\t }{1-\frac{a^2}{\ell^2}  \cos^2\t }, \\
    & b(\t)=\frac{\delta}{V\a}\frac{ r_+^2-a^2 \cos ^2\theta }{ r_+^2+a^2 \cos ^2\theta }.   
   \end{align}
 where $r_+, V, \delta, a, \alpha,$ are, for now, arbitrary constants.
   These are supplemented by the extremality conditions which, in the context of Sen's entropy function formalism arise from Eqs. \eqref{Eq1KNAdS}-\eqref{Eq2KNAdS}: 
   \begin{align}
   & \alpha ^2 \lambda -\frac{4 a^2 r_+^2}{V^2}=0, \label{Eq3:KN-AdS} \\
& 4 a^2 (\ell-r_+) (\ell+r_+)+\ell^2 \left(\delta ^2 e^2-4 r_+^2\right)-12 r_+^4=0
   \end{align}
Note that we can recover the extremality relations of the previous section by setting $\delta=0$. Through the regularity condition \eqref{lambdacondition}, we  find
  \begin{equation}
  \label{Eq:LambdaKerr-AdS}
\lambda=\frac{\left(r_+^2+a^2 \right)^2}{\left(1-\frac{a^2}{\ell^2} \right)^2}. 
\end{equation}
which is similar to the expression in the previous section, except that dependence on the electric field  enters through $r_+$. \par 
Now we construct the entropy function
\begin{equation}
F=2\pi \left(J\a+qe-2\pi \int_0^\pi d\t \mathcal{L}_{EMGB} \right).
\end{equation}
The constants $J$ and $q$ can be determined from $\frac{\partial F}{\partial\a}=0, \frac{\partial F}{\partial e}=0$. They are given by: 
\begin{align}\label{Eq:Implicit}
& J=\frac{8 \pi  a^2 \delta ^2 e^2 r_+^2 (a-r_+) (a+r_+)}{\alpha ^3 \sqrt{\lambda } V^3 \left(a^2+r_+^2\right)^2}+\frac{2 \pi  \alpha  \lambda ^{3/2} V \left(\ell^2+3 r_+^2\right)}{a^2 \ell^2 r_+^2}\\
& q=-\frac{2 \pi  \delta ^2 e (a-r_+) (a+r_+) \left(4 a^2 r_+^2+\alpha ^2 \lambda  V^2\right)}{\alpha ^2 \sqrt{\lambda } V^3 \left(a^2+r_+^2\right)^2}. 
\end{align}
Inserting these into the entropy function we find
\begin{equation}
S_{Sen}=16\pi^2 \sqrt{\lambda}+64\pi^2 \g = 16\pi^2 \frac{\left(r_+^2+a^2 \right)}{1-\frac{a^2}{\ell^2}} +64\pi^2 \g. \label{SBHKNAdS}
\end{equation}
The entropy should, of course,  be expressed as $S(J,q)$. The current expression is  implicit but can, in principle, be inverted using Eq. \ref{Eq:Implicit}. The structure of the entropy,  again, matches our general formula \eqref{Bhcorrected}.

\subsection{Agreement with AdS$_4$ black holes in gauged supergravity}\label{SubSec:SugraAdS4}

In this section we compare the results obtained using Sen's entropy function formalism and the symmetries of the near-horizon geometry to those obtained from the full  of Kerr-AdS and Kerr-Newman-AdS black holes.

\subsubsection{Kerr-AdS blackholes}
The Kerr-AdS black holes in the Boyer-Lindquist coordinates can be written as \cite{carter1968,Dias:2012pp}: 
\be
  ds^2 = - \frac{\Delta_r}{W} \left(dt - \frac{a\, \textrm{sin}^2 \theta}{\Xi} d\phi \right)^2 + W \left(\frac{dr^2}{\Delta_r} + \frac{d\theta^2}{\Delta_\theta} \right)^2 + \frac{\Delta_\theta\, \textrm{sin}^2 \theta}{W} \left[a\, dt - \frac{r^2 + a^2}{\Xi} d\phi \right]^2\, ,\label{eq:AdS4Metric2}
\ee
where 
\begin{eqnarray}
\Delta_r (r)&=& (r^2+a^2)\left(1+\frac{r^2}{\ell^2} \right)-2 mr, \nonumber \\
  \Delta_\theta & \equiv &1 - \frac{a^2}{\ell^2}\, \textrm{cos}^2 \theta\, , \label{Eq:Deltatheta}\\
  W & \equiv& r^2 + a^2\, \textrm{cos}^2 \theta\, ,\\
  \Xi & \equiv &1 - \frac{a^2}{\ell^2}\, , \label{Eq:Xi}
\end{eqnarray}
 In the extremal limit, we have 
\begin{equation}
\Delta_r(r_+)=0, \ \Delta_r'(r_+)=0
\end{equation}
where $'$ denotes derivative with respect to $r$. Therefore, near $r=r_+$ we can expand: 
\begin{equation}
 \Delta_r (r)=V (r-r_+)^2+\mathcal{O}(r-r_+)^3. 
 \end{equation} 
 The Bardeen-Horowitz \cite{Bardeen:1999px} limit is taken by \cite{Lu:2008jk,Chow:2008dp}  
 \begin{equation}
 r=r_+(1+\epsilon y), \ \phi=\varphi+\frac{a\Xi}{r_+^2+a^2}t, \ t=\frac{r_+^2+a^2}{\epsilon r_+ V}\tau . \label{Bardeen-Horowitz}
 \end{equation}
After performing these transformations and taking the $\epsilon\to 0$ limit we obtain the near horizon geometry: 
\begin{equation}
ds^2_{NH}=\frac{W_+}{V} \left( -y^2 d\tau^2+\frac{dy^2}{y^2}+\frac{V}{\Delta_\theta} d\theta^2 \right)+\frac{\Delta_\theta \left(r_+^2+a^2 \right)^2 }{W_+ \Xi^2} \sin^2\theta \left(d\varphi +\frac{2 r_+ a \Xi}{V(r_+^2+a^2)}yd\tau \right)^2 \label{NHmetric1}
\end{equation}
where we have defined: 
\begin{equation}
W_+=r_+^2+a^2 \cos^2\theta . 
\end{equation}
Note that this is of the same form of our ansatz \eqref{metricKerr-AdS} and from this we can extract the relevant functions which match with Section \ref{SubSec:AdSRotGB}. More precisely, we have $\beta(\theta)=W_+/\Delta_\theta$ and $\lambda = (r_+^2+a^2)^2/(1-a^2/\ell^2)^2$ just like in equations \ref{Eq:betaKerr-AdS} and \ref{Eq:LambdaKerr-AdS} from  Section \ref{SubSec:AdSRotGB}, respectively.  Note, in particular, that we recover the Bekenstein-Hawking entropy precisely. As a bonus, Sen's entropy function formalism provides a shortcut to computing the correction to the entropy due to higher-derivative corrections which we evaluated in Section \ref{SubSec:AdSRotGB}.\par

\subsubsection{The Kerr-Newman-AdS black hole}
In the case of Kerr-Newman-AdS black holes the metric in Boyer-Lindquist coordinates is given by \cite{Caldarelli:1999xj}: 
\be
  ds^2 = - \frac{\Delta_r}{W} \left(dt - \frac{a\, \textrm{sin}^2 \theta}{\Xi} d\phi \right)^2 + W \left(\frac{dr^2}{\Delta_r} + \frac{d\theta^2}{\Delta_\theta} \right)^2 + \frac{\Delta_\theta\, \textrm{sin}^2 \theta}{W} \left[a\, dt - \frac{r^2 + a^2}{\Xi} d\phi \right]^2\, ,\label{eq:AdS4Metric3}
\ee
where all functions are defined previously in equations (\ref{Eq:Deltatheta}) -- (\ref{Eq:Xi}), except for $\Delta_r $ which  is now given by:
\begin{equation}
\Delta_r (r)= (r^2+a^2)\left(1+\frac{r^2}{\ell^2} \right)-2 mr +q^2.  
\end{equation}
In the extremal limit, we again have: 
\begin{equation}
\Delta_r(r_+)=0,\  \Delta_r'(r_+)=0. 
\end{equation}
The Bardeen-Horowitz limit is taken exactly in the same way as before. Therefore Eq. \eqref{NHmetric1} is also the near horizon metric in this case. \par 
We also have non-trivial gauge field which is given by  \cite{Caldarelli:1999xj}: 
\begin{equation}
A=-\frac{qr}{W^2}\left(dt-\frac{a\sin^2\theta}{\Xi}d\phi \right). 
\end{equation}
After taking the Bardeen-Horowitz limit \eqref{Bardeen-Horowitz} we can find the field-strength tensor as: 
\begin{equation}
F_{NH}=-\frac{q}{V}\frac{r_+^2-a^2\cos^2\theta}{r_+^+-a^2\cos^2\theta}d\tau\wedge dy+\frac{qar_+}{W_+^4}\sin (2 \theta)d\theta \wedge \left( \frac{r_+^2+a^2}{\Xi}d\varphi +\frac{2ar_+}{V} y d\tau \right). 
\end{equation}
This is in agreement with section \ref{KNAdS}.

\subsubsection{The supersymmetric embeddings and global aspects}

Let us finally discuss an important point in confronting our analysis with the more traditional approach to the black hole entropy which starts with the full  gravity solution. Let us phrase the question in the context of the most general AdS$_4$ black holes that have been the object of much recent work related to the microscopic discussions of the entropy. Recall that in the context of  $4d$ $\mathcal{N}=4$ gauged supergravity with gauge group $U(1)\times U(1)$  explicit solutions are known describing non-extremal black holes \cite{Chong:2004na}. The supersymmetric limit and important global aspects of the solution were originally presented  in \cite{Cvetic:2005zi} and  later discussed in \cite{Chow:2013gba} where a few typos in the original work were corrected. More recently,  vis-\`a-vis microscopic entropy discussions, two very lucid accounts were given  in \cite{Choi:2018fdc} and \cite{Cassani:2019mms}.

These black holes are conveniently described by four parameters $(m,a,\delta_1, \delta_2)$ which we can roughly think of proxies for the mass, angular momenta and two electric charges.  In particular \cite{Chong:2004na,Cvetic:2005zi}:

\begin{eqnarray}
E&=&\frac{m}{2G \Xi^2}(\cosh 2\delta_1 +\cosh2\delta_2), \quad J=\frac{m \,a }{2G \Xi^2}(\cosh 2\delta_1+\cosh 2\delta_2), \nonumber \\
Q_1&=& \frac{m}{4G\Xi}\sinh 2\delta_1,  \qquad \quad \qquad Q_2=\frac{m}{4G \Xi}\sinh 2\delta_2,
\end{eqnarray}
where $\Xi$ is as in equation \ref{Eq:Xi}. When specializing to the supersymmetric case, one needs to impose the following relation \cite{Cvetic:2005zi}:
\be
e^{2\delta_1+2\delta_2}=1+2\frac{ \ell}{a}.
\ee
This relation naturally reduces the four independent parameters to three. A further global condition was introduced in \cite{Cvetic:2005zi} to guarantee the existence of a regular horizon (see also  \cite{Choi:2018fdc} and \cite{Cassani:2019mms}):
\be
\left(\frac{m}{\ell}\right)^2 = \frac{\cosh^2(\delta_1+\delta_2)}{e^{\delta_1+\delta_2}\sinh^3(\delta_1+\delta_2)\sinh 2\delta_1 \sinh 2\delta_2}.
\ee
As a result of this global constraint, the solution ends up depending on two free parameters.  For example, the entropy takes the form
\be
S=\frac{2\pi \ell^2}{G(e^{2\delta_1+2\delta_2}-3)}.
\ee
More relevant for our context is that the extra global condition implies a relationship among the charges. Through the  BPS condition the energy is related to the other charges as $E=2Q_1+2Q_2+J/\ell$. One then finds that the global condition implies that the angular momentum is related to the electric charges according to\cite{Choi:2018fdc,Cassani:2019mms}
\be
J=(Q_1+Q_2)\ell \left(\sqrt{64G^2Q_1Q_2/\ell^2 +1}-1\right).
\ee
The microscopic entropy of AdS$_4$ black holes has been  provided in \cite{Choi:2019zpz} and \cite{Nian:2019pxj} by considering the dual field theory side; in that context the constraint appears as a condition for the entropy to be real. 

In previous sections we  considered the case of one electric charge which corresponds to the situation with  $\delta_1 =\delta_2$ whose supergravity solution was  constructed in \cite{Kostelecky:1995ei} (see also \cite{Caldarelli:1998hg}).  Since this type of black holes is  more universal in their higher dimensional embedding, microscopic foundations for the entropy can also be provided using M5 branes, as was the case in  \cite{Bobev:2019zmz,Benini:2019dyp}; the effect of the charge constrains was also captured in that general situation.

What is relevant for our study of Sen's classical entropy formalism is that  we are unable to access the relation between the angular momentum and electric charged in this solution, $(J,Q)$.   Therefore, Sen's classical entropy function formalism seems to miss important global constraints on the solution by focusing on its near horizon geometry.

\section{Conclusions and Outlook}\label{Sec:Conclusions}

In this manuscript we have explored Sen's entropy function formalism in the context of asymptotically AdS$_4$ black holes. In section \ref{Sec:Static} we discussed static black holes and  demonstrated that the entropy arising from Sen's classical entropy function formalism matches the Wald entropy computation in four-dimensional supergravity theories with higher-derivative terms incorporated with guidance from conformal supergravity \cite{Bobev:2020egg}. In Section \ref{Sec:Rotating} we explored rotating black holes and showed that Sen's entropy function formalism leads to a near-horizon geometry that is sufficient for the computation of corrections to the entropy due to higher-derivative terms. Moreover, the background solution obtained using only knowledge of the near-horizon symmetries coincides with the   Bardeen-Horowitz near-horizon limit of rotating, electrically charged AdS$_4$ black holes already known in the literature by solving the full Einstein equations.  These examples demonstrate that Sen's entropy function formalism is a highly efficient approach to incorporating higher-derivative corrections to the entropy of AdS$_4$ black holes albeit missing certain global aspects of the full solution. 

There are number of interesting directions that we think would be fruitful to explore in the future. The most immediate one is exploring Sen's entropy function formalism in other dimensions, for example, in asymptotically AdS$_{5,6,7}$.  Previous investigations of Sen's entropy function formalism for static AdS$_{4,5}$ black holes \cite{Morales:2006gm} confirmed its efficacy. The really new direction would be to explore rotating black holes for which there is now a  microscopic understanding in terms of the superconformal indices in the corresponding dual field theories. Our work in section \ref{Sec:Rotating} is just the first step in this direction which promises to be quite rich in higher dimensions where multiple angular momenta are possible.

Given that  Sen's entropy function formalism relies on knowing only the near-horizon geometry, one would expect that it will be sufficient to understand the entropy, including higher derivative corrections, of arbitrary supersymmetric AdS$_4$ black holes. This power becomes crucial in the face of what we call {\it missing black hole solutions}.  Recent important developments in supersymmetric localization essentially predict the entropy of the dual  black holes asymptotically to AdS$_4\times M^7$ and  AdS$_5\times M^5$ for large classes of Sasaki-Einstein manifolds  $M^7$ and $M^5$; some of these results are partially supported in supergravity \cite{Hosseini:2019ddy,Gauntlett:2019roi,Kim:2019umc}. Explicitly constructing those black holes is an ongoing challenge in the supergravity community. Sen's entropy function formalism might provide enough near-horizon geometry information to start filling this gap.

Finally, it would be interesting to revisit Sen's {\it quantum} entropy function formalism for AdS black holes. The quantum entropy function formalism, as described for asymptotically flat spacetimes  in \cite{Sen:2008vm} allowed to understand quantum corrections   to the entropy  of various string theory black holes  \cite{Sen:2007qy} exploiting an AdS$_2$/CFT$_1$ intuition  \cite{Sen:2008yk}. The first attempts to apply Sen's quantum entropy function formalism to AdS$_4$ black holes in  \cite{Liu:2017vll,Jeon:2017aif} did not match the field theory prediction. Various computations have indicated that knowledge of the near-horizon geometry is not sufficient to reproduce certain field theory predictions and that agreement is found when focusing on the AdS$_4$ asymptotics  \cite{Liu:2017vbl,Gang:2019uay,Benini:2019dyp,PandoZayas:2020iqr}.    It would be quite interesting to understand if there is some connection between the near-horizon and asymptotic computation that could be explained in terms of extra hairy degrees of freedom in AdS.

\section*{Acknowledgements}
We are thankful to Sangmin Choi, Chandramouli Chowdhury, Marina David and  Jun Nian. We are particularly grateful to  Jos\'e  Francisco Morales for various discussions and clarifications regarding his previous work on this topic  \cite{Morales:2006gm}. We also thank  Nikolay Bobev, Pieter Bomans and especially  Valentin Reys for clarifying related technical details. JKG acknowledges the postdoctoral program at ICTS for funding support through the Department of Atomic Energy, Government of India,
under project no. 12-R\& D-TFR-5.10-1100.  LPZ's work is  supported in part by the U.S. Department of Energy under grant de-sc0007859.

\bibliographystyle{JHEP}
\bibliography{Wald-Entropy}

\end{document}